\documentclass[10pt,preprint2]{aastex}

\usepackage{geometry}
\usepackage{times}
\begin{document}
\title{On The Size of Structures in the Solar Corona}

\author{C.E. DeForest}
\affil{Southwest Research Institute}

\affil{\center{\emph{[In press, \apj - est. 2007 June]}}}

\begin{abstract}
Fine-scale structure in the corona appears not to be well resolved by
current imaging instruments. Assuming this to be true offers a simple
geometric explanation for several current puzzles in coronal physics,
including: the apparent uniform cross-section of bright threadlike
structures in the corona; the low EUV contrast (long apparent scale
height) between the top and bottom of active region loops; and the
inconsistency between loop densities derived by spectral and
photometric means.  Treating coronal loops as a mixture of diffuse
background and very dense, unresolved filamentary structures address
these problems with a combination of high plasma density within the
structures, which greatly increases the emissivity of the structures,
and geometric effects that attenuate the apparent brightness of the
feature at low altitudes.  It also suggests a possible explanation for
both the surprisingly high contrast of EUV coronal loops against the
coronal background, and the uniform ``typical'' height of the bright
portion of the corona (about 0.3 $R_{\odot}$) in full-disk EUV images.
Some ramifications of this picture are discussed, including an
estimate (10-100 km) of the fundamental scale of strong heating events
in the corona.
\end{abstract}


\section{Introduction\label{sec:intro}}

\begin{figure}[!tb]
\center{\includegraphics[
  width=3.2in,
  keepaspectratio]{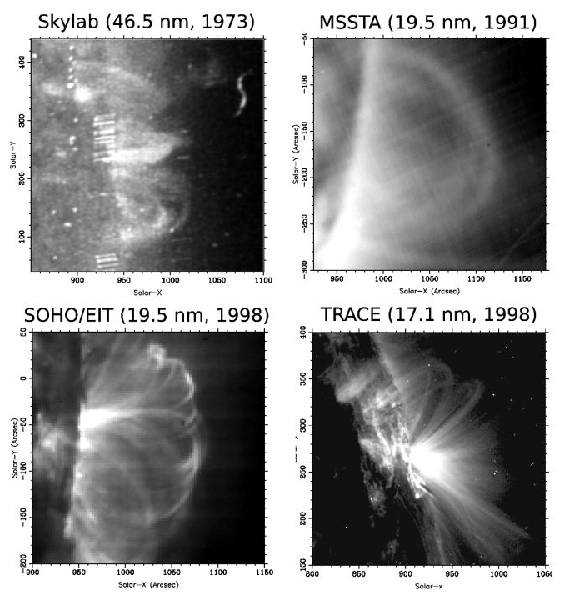}}
\caption{\label{fig:sample}Limb images at the same spatial scale from four
different UV imagers demonstrate the long-observed constant-width
appearance of coronal loops. The apparent constant width varies from
instrument to instrument.  It is due to resolution effects in data
from the Skylab ``overlappograph'' (\citealt{Tousey1977,Feldman1987})
at \textasciitilde{}8'', the Multi-Spectral Solar Telescope Array
(\citealt{Walker1991}) at \textasciitilde{}10'', and the Extreme-Ultraviolet
Imaging Telescope (\citet{Delaboudiniere1995}) at 5''. I argue that
the Transition Region and Coronal Explorer (\citealt{Handy1998}),
the highest resolution existing EUV imager at \textasciitilde{}1'',
also does not resolve the filamentary structure within active regions.}
\end{figure}

Since the introduction of EUV coronal imaging, bright coronal loops
have been seen to have uniform thickness; several examples are
illustrated in Figure \ref{fig:sample}. Because coronal bright
structures are thought to be magnetically confined, this is a
surprising result.  Uniform-thickness flux bundles require constant
magnetic field strength along the feature; this implies that tall,
uniform-thickness bright structures in the corona must somehow be
confined not only against gas pressure, but also against magnetic
pressure.  Much recent work has been devoted to understanding the
physics of these structures, and several terms have been used to
describe them.  I suggest ``thread'' as a purely observational term to
refer to thin, curvilinear features such as may be seen in the image
plane of a solar instrument, avoiding any particular physical model
for the corresponding coronal structure; phrases such as ``filamentary
structure'', ``flux tube'', or ``elementary structure'', which
themselves carry some implicit meaning about the physics, may then be
used to describe the structure in the corona itself.  Furthermore,
throughout this article I refer to objects in the image plane of a
telescope as ``features'' and objects in the solar corona as
``structures''.

Thin thread features are nearly always seen embedded in larger coronal
features (for example active region loops) that expand with altitude
as expected from force-free or potential field lines, but the smallest
threads very clearly have uniform thickness in solar EUV images.
Considerable effort has been put into understanding the physics of the
corresponding solar structures.  \citet{AschwandenNitta2000} pointed
out that inhomogeneous structure can influence the ionization
temperature gradient inferred via filter-ratio or differential
emission-measure techniques applied to EUV image features.  Comparison
of \emph{TRACE}-visible active region loops to simple hydrostatic
atmospheric models shows that the tops of tall threaded loops seem
overdense by a factor of up to 100 compared to hydrostatic solutions
or, equivalently, that the threads have apparent intensity scale
heights much longer than the calculated thermal scale height in the
corona
(e.g. \citealt{Doyle1985,AschwandenNitta2000,WinebargerWarrenMariska2003,Fuentes2006}).
Supporting a longer-than-thermal density scale height over such ranges
requires extreme measures, such as ballistic siphon flows or wave
pressure, that are not reflected in the spectral data. Simple
calculation shows that doubling the density scale height of a 0.2
$R_{\odot}$ tall structure requires a basal speed of over 150 km
sec$^{-1}$, compared to typical active region loop Doppler shifts of
only 20-100 km sec$^{-1}$ as observed with \emph{SOHO}/CDS
(\citealt{Fredvik2002}).  Taller structures or more uniform density
profiles require even higher speeds for such support.

The emission from some of these filamentary structures has been
identified by \citet{Aschwanden2005b} as nearly thermally homogeneous,
leading those authors to identify some particularly cleanly presented
threads as {}``elementary'' coronal structures having uniform
cross-section versus position and a filling factor nea unity with
constant temperature and density.  \citet{Fuentes2006} have
made detailed quantitative studies of threads' profiles in a large
sample of active regions . However, similar thin, bright active region
loops yield much higher densities from density-sensitive line ratios
than expected from photometric considerations
(\citealt{WarrenWinebarger2003}).

These inconsistencies all appear to be tied to a measurement that is
not solid: the width of the structures themselves.  The uniform
thickness features seen in each panel of Figure \ref{fig:sample} are
close to the resolution limit of the corresponding telescope.  In
fact, \emph{TRACE}, the highest resolution EUV telescope currently
available, shows that active region loops appear composed of
uniform-thickness threads about 1-2 Mm in width, which diverge one
from the other as might be expected of individual field lines.  One
may conclude that the constant width of threads observed with earlier
instruments was an instrumental effect: the underlying structures must
have expanded with altitude as do complete bundles of threads in \emph{TRACE}
images.  That conclusion, in turn, raises the question of whether
constant-width features in the \emph{TRACE} images are due to thin,
variable width structures that are simply not well resolved.

Assuming that threaded loops observed with \emph{TRACE} are not
fully resolved eliminates two important problems. First, it avoids the
theoretical problem of how to keep the structures confined compared to
the surrounding general expansion, because if unresolved the
structures do not have to be confined at all compared to a force-free
field.  In that case, they may expand laterally with altitude at the
same relative rate as a force-free flux tube, provided only that each
feature's cross-section remains below or at least near the telescope
resolution limit. Secondly, it provides a geometric explanation for
the very long apparent scale height of active region loops seen with
\emph{TRACE}, rendering the measured intensity profiles consistent
with hydrostatic equilibrium within each loop: if the size of the
feature is permitted to vary with height, then geometrical
considerations can provide sufficient extra brightness at high
altitudes to explain the typical coronal intensity profiles, even if
the loops are supported only hydrostatically.

Coronal loops are visible to surprisingly high altitudes with EUV
imagers in general and \emph{TRACE} in particular. It is not uncommon
for an active region on the limb to be seen out to 140 Mm (0.2 $R_{\odot}$)
or higher, and the ``background corona'' is visible out to about 0.3
$R_{\odot}$ in typical images from EIT \citealt{Delaboudiniere1995}. The
scale height of 1-2 MK plasma near the solar surface is 50-100 Mm
respectively (0.07 - 0.14 $R_{\odot}$), smaller than these typical EUV
feature heights.  If the plasma in the loops is in hydrostatic
equilibrium, both the top and bottom of a uniform-thickness coronal
structure as tall as 0.2-0.3 $R_{\odot}$ should not be readily visible in a
linear scale image.  However, if the thickness of the structure varies
then the instrument images a different volume of plasma in a single
pixel at different altitudes. Higher up, the tube should be both
broader and thicker, and hence brighter, than would be expected for a
uniform thickness flux tube.  This effect is large enough to
compensate for a hydrostatic density profile with altitude, out to
(coincidentally) 0.2-0.3 $R_{\odot}$, after which the hydrostatic lapse
rate dominates and the features rapidly dim.  In turn, this provides a
very clean explanation for the overall coronal morphology as seen with
EUV imagers.

\begin{figure*}[!tb]
\center{\includegraphics[%
 width=4.5in,
 keepaspectratio]{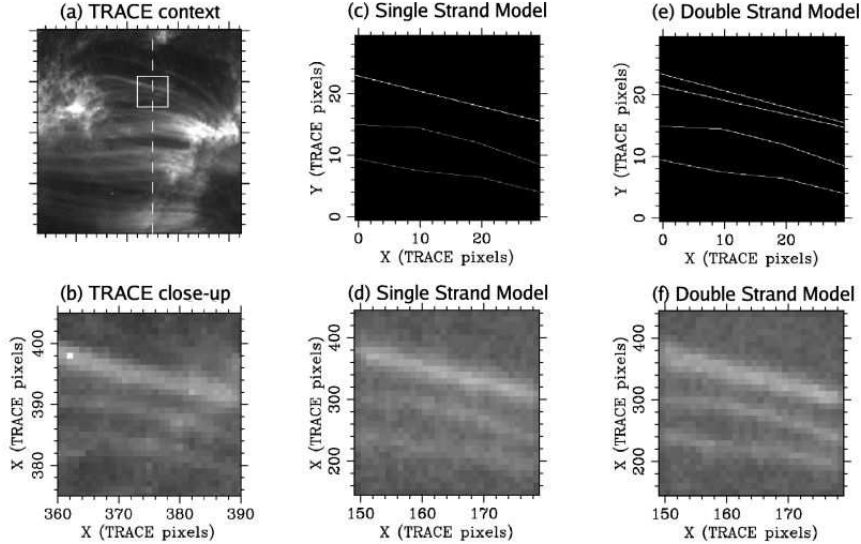}}
\caption{\label{fig:forward_model}
A simple forward model of threaded loops in an active region imaged
by \emph{TRACE} in the 171 \AA\ passband on 1999 April 7 at 1:26 UT.  (a) 
Context image of the active region.  (b) Close-up of a small region
of interest.  (c) A simple model of ``elementary'' coronal structures
(d) The model after convolution, pixelation, noise and pedestal were added.
(e) A tapered dual-strand model of the top thread.  (f) Dual-strand model
after convolution, pixelation, noise, and pedestal.  Neither of the 
two models is readily distinguished from the \emph{TRACE} image.
}
\end{figure*}

In the following sections, I explore the hypothesis that elementary
coronal features are \emph{not} resolved and demonstrate that it is
consistent with individual \emph{TRACE} images and a hydrostatic
corona. \S\ref{sec:resolution} is a discussion of telescope
resolution, including a forward model of a simple structure close to
the resolution limit of a telescope, and a demonstration of confusion
effects in interpretation of solar data. \S\ref{sec:profiles}
demonstrates that unresolved structure can greatly enhance the
apparent scale height of bright features in the
corona. \S\ref{sec:model} uses a forward model to reproduce the
general appearance of active regions seen with \emph{TRACE} using only
geometry and hydrostatic density profiles.  \S\ref{sec:trace} is a
study of the morphology of several threaded active regions observed
with \emph{TRACE}, including estimates of the size scale of unresolved
features at the base of the active regions.  All of these analyses are
simple but sufficient to demonstrate a geometrical effect that has
previously been ignored and that strongly affects interpretation of
coronal EUV images. In \S\ref{sec:discussion}, I conclude that
unresolved loop structure is the simplest explanation for the observed
intensity profile and apparent uniform thickness of active region
loops observed with \emph{TRACE}, discuss implications for the time
evolution of active region threads, and call for higher resolution
observations.

\section{Telescope resolution\label{sec:resolution}}

The \emph{TRACE} telescope has a pixel size of 0.5 arc seconds
(\citealt{Handy1998}) and a point-spread function with a full-width at
half-maximum of about 2.25 pixels
(e.g. \citealt{Golub1999,Gburek2006}).  This is similar to the
observed width of coronal threads in \emph{TRACE} EUV images of active
regions, weakening inferences that may be drawn from the image
geometry about the size of the corresponding coronal
structures on the Sun.  Some authors (e.g. \citealt{Fuentes2006}) have taken
great care in analyzing the size of active region filamentary
structures based on observations of threads, but even such careful
studies may have doubt shed on them by interaction between barely
resolved structures, background noise, and other superposed features
in the image plane.  These effects can mimic those
of a larger PSF, greatly weakening the commonly-held hypothesis
that threads are adequately resolved in \emph{TRACE} images (and hence that
the size of solar filamentary structures may be determined from the
image-plane size of the corresponding features).

To demonstrate the effects of random noise and telescope PSF on
simple, compact linear structures, I implemented a simple 2-D forward
model of the \emph{TRACE} point-spread function to demonstrate the effects of
noise and background structure on resolution of fine-scale loops.
Figure \ref{fig:forward_model} shows a simple treatment of three
threads in an active region observed by \emph{TRACE}.  A small
region-of-interest was modeled as three or four infinitely thin
threads that were convolved with a circularly symmetric model PSF with
a FWHM of 2.25 \emph{TRACE} pixels; pixelated; and subjected to background
pedestal and both uncorrelated (``photon'') and locally correlated
(``solar background'') noise fields.  I ran two forward models of the
image: one with a single bright thread at the top and one with two
less bright threads at distances of up to two \emph{TRACE} pixels.  The
models were not distinguishable, indicating that structures with
apparent size less than two \emph{TRACE} pixels cannot be distinguished
visually from structures of zero width.  In either case, the resulting
bright features had apparent visual widths of 5-6 \emph{TRACE} pixels, or
about 2-3 PSF widths, in the presence of background photon noise and
solar structure.  This highlights the difficulty
of estimating feature widths with a non-Gaussian PSF (as found by 
\citealt{Gburek2006}) in the presence of background noise:  the visual
width or even the measured FWHM of an image plane feature in the presence
of additive background signal and photon noise may be significantly wider
than expected from straightforward analysis of the PSF size.

\begin{figure*}[!tb]
\center{\includegraphics[%
 width=5.75in,
 keepaspectratio]{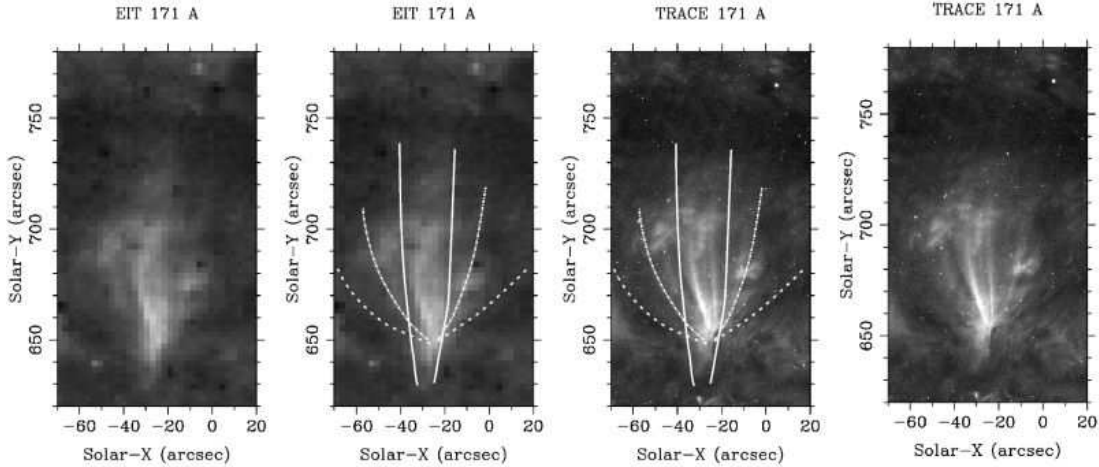}}
\caption{\label{fig:res_comparison}
Two views of the same polar plume, imaged by \emph{TRACE} and \emph{SOHO}/EIT at
1999 Aug 27 19:00 UT, show the interplay between instrument resolution and derived
structure size, even in features many pixels across.  In the central panels, visual
edges of the plume have been traced using the EIT data (solid lines), using the \emph{TRACE}
data with a conservative approach (dot-dashed lines), and using the \emph{TRACE} data with
the faintest visible edge at the extreme periphery of the plume (dashed lines).  The 
outlines of the plume are surprisingly dependent on the instrument.
}
\end{figure*}

The visual result here is consistent with the results of
\citet{Fuentes2006} in analyzing forward models of cylindrical
structures near the \emph{TRACE} resolution limit.  Both results
highlight the difficulty of separating the size of structures near the
telescope's ultimate limit.  The Fuentes analysis uses the standard
deviation of the feature's brightness considered as a random-variable
distribution, and is more rigorous than the present visual-width
argument, but also demonstrates the importance of telescope PSF to the
derived width of small structures in the corona from features in the
image plane.  Fuentes et al. argue that the PSF is small enough to
not affect their main results on thread width versus altitude, but
that argument depends strongly on the derived PSF of the instrument
and may depend on other image interpretion effects that are not
considered directly in that analysis.

In real data several effects worsen considerably the strength of image
interpretation beyond a simple linear analysis of feature spreading by
convolution with a PSF: most solar structures are not compact in
cross-section; the image background contains features that can be
confused with the feature of interest; and brightness variation across
the cross-section of a linear feature may interact with background
noise to produces spurious feature ``edges'' that are inconsistent
with edges derived from the same structure viewed at a different
resolution.

Figure \ref{fig:res_comparison} shows an example of the importance of
resolution even to interpretation of structures that are significantly
larger than the point-spread function of the observing telescope.  Two
images are shown of the same polar plume, taken with
\textasciitilde{}5 arc second resolution (by EIT) and with
\textasciitilde{}1 arc second resolution (by \emph{TRACE}).  The two images
were collected within 30 seconds of one another and were both in the
171 \AA\ EUV band, so the strongest differences between them are due to
the instruments themselves.  While some dark pixels are visible in the
EIT image, and both images are subject to different patterns of cosmic ray
hits, the largest difference is in the resolution of the images.

The EIT image shows the plume as a bright, thin feature embedded in
a larger, round diffuse region on the Sun.  With the higher resolution
available to \emph{TRACE}, we can identify many bright features in the
plume and recognize that the plume has both a diffuse component and a
threadlike component.  By comparing resolution effects between the
images at 5 arc second resolution and 1 arc second resolution near the
limits of EIT, we can identify the sorts of visual effects that are to
be expected near the resolution limit of \emph{TRACE}.

The outer panels of Figure \ref{fig:res_comparison} are raw images;
the inner panels are the same two images, but with co-aligned visually
identified edges marked.  The solid lines in the central panels show
edge tracings using the EIT data alone; the dashed lines show edge
tracings using the \emph{TRACE} data, with the inner (dot-dash) set
corresponding to the clearest visible edge of the plume as a whole and
the outer (dash) set corresponding to the outermost visually
identifiable edge.

The first, and most relevant, discrepancy to notice between the
instruments is that the identified expansion factor is radically
different between the two data sets.  Although the plume is
\textasciitilde{}6 EIT pixels across at its brightest position
(considerably larger than the EIT point-spread function) the expansion
factor of the EIT edge tracing is less than two in the first 30 Mm of
altitude.  In the \emph{TRACE} image there is both a diffuse component
to the plume and also many small, bright threads that outline a much
broader magnetic structure with a much larger overall expansion ratio than 
is visible with EIT.

The discrepancy in expansion ratio across the two instruments can be
explained in terms of the interaction between resolution, geometric
cross section, and background features.  Near the center, the plume is
denser and has a higher depth along the line-of-sight, making the core
more visible than the exterior and giving rise to a visual edge with
low expansion factor in the EIT image.  At the base of the plume,
resolution effects limit the smallest visible size of the overall
structure, and higher up, only the brightest portion of the plume is
visible and separable from the solar background, reducing the apparent
cross section of the structure.  These two effects conspire to produce
a dramatically lower expansion ratio than is apparent in the
\emph{TRACE} data.  Note that near the base of the plume the EIT visual
edges cut laterally across field lines: the edge is defined by the
gradient in intensity, which in turn depends both on variation in the
visual chord length and also on the falling density versus altitude.
The brightness at the base of the plume includes components both from
the individual threads that are visible in the \emph{TRACE} image and also
from the diffuse material that fills the bulk of the plume's volume;
but even 30-40 arc seconds above the base of the plume, the diffuse
emission falls below the intensity of background features and the 
threads dominate the visual appearance of the plume.

Other types of confusion are worth mention.  Surrounding the base of
the plume in the EIT image is a large, roughly circular region of
diffuse emission that appears to be a supergranule or part of the
network.  Comparison with \emph{TRACE} indicates that, although network
brightenings are evident in the background, much of the brightness
around the plume core is in fact due to the plume itself.

Additional problems with the lower resolution image include a lack of
visual distinction between foreground and background objects.  This leads
both to an apparent twisting visible as a crescent-shaped bright feature in
the plume core, and to an error in the position of the plume
footpoint.  At higher resolution the crescent is resolved as a
coincidental alignment of a background network brightening and a
foreground thread, while the base of the plume is seen to be confused
with a foreground network brightening that extends the visual feature
in the EIT image.

\begin{figure}[!tb]
\center{\includegraphics[%
  width=0.93\columnwidth]{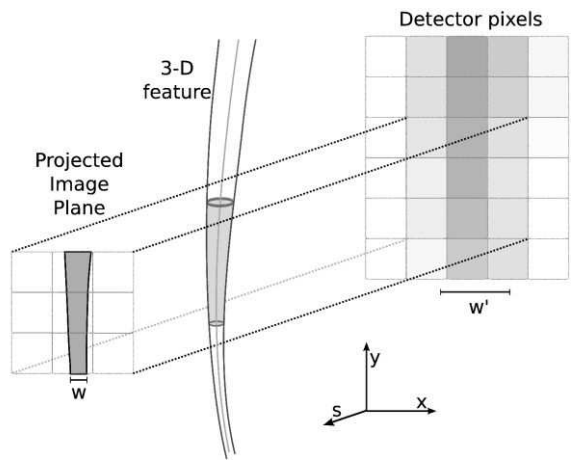}}
\caption{\label{fig:geometry}Geometry of optically thin solar imaging with
finite resolution and pixel size. Features exist in 3-D (center) but
images are brightness integrals along the line of sight $s$ (left).
Physical instruments have instrumental spreading and pixel quantization
(right), so that a feature with image plane width $w$ in fact yields
a feature of apparent width $w'$. The apparent brightness of the
feature is proportional both to the emissivity within the feature
and to the volume (shaded at center) of the feature within each pixel.}
\end{figure}

\begin{figure*}[t]
\center{\includegraphics[%
  width=0.6\paperwidth]{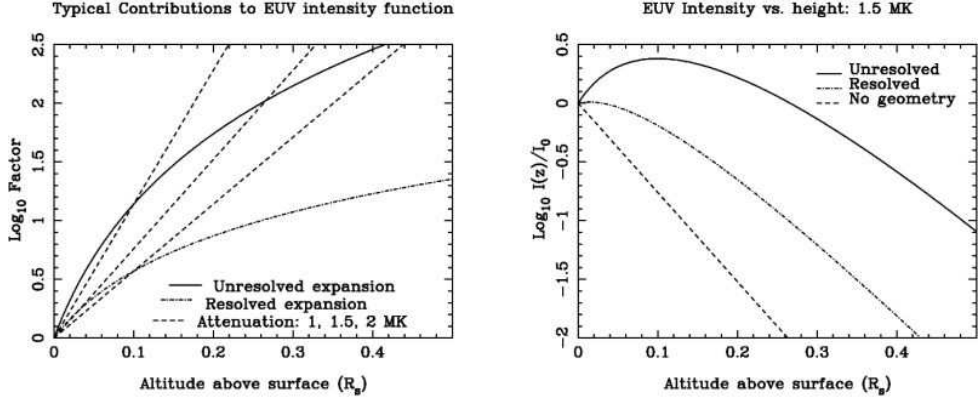}}
\caption{\label{fig:Comparison of exponential and parabolic growth}Unresolved,
expanding filamentary structures explain the strong visibility of
coronal features in EUV, out to about$1.3\, R_{\odot}$. LEFT: interplay of
expansion for a typical active-region and three different scale heights (1.0
MK, 1.5 MK, and 2.0 MK).  RIGHT: relative intensity versus height for
structures at 1.5 MK, both with and without geometrical corrections.
Without the geometric correction, coronal features are expected to be
strongly attenuated above about 0.1 $R_{\odot}$ for all but the hottest
features; with the correction, unresolved or barely-resolved features
can be expected to remain visible out to \textasciitilde 0.3 $R_{\odot}$ (as is seen
with EIT and \emph{TRACE}). }
\end{figure*}

The presence of various ambiguities and confusion in even quite large
structures indicates that one must use extreme caution when analyzing
structures that are close in size to the resolution of the source
telescope, even when the apparent structure is several times larger
than the PSF of the instrument.  The reason is that visual context is
needed to distinguish structural elements that give rise to the
features in an image plane.  Near the resolution limit, textural and
alignment cues are not available because such cues are blurred out
by the finite resolving power of the instrument.

In this case, the width of the EIT-determined plume structure is
\textasciitilde{}6 EIT pixels (15 arc seconds) near the base at
solar-Y=660, quite a bit larger than the EIT PSF width of
\textasciitilde{}1-2 pixel (\citet{Delaboudiniere1995}); but improving
the resolution by a factor of 3-5 (in the \emph{TRACE} image) reveals
the plume to be much wider: 30-60 arc seconds across at the same altitude.

\section{Feature intensity profiles\label{sec:profiles}}

EUV images from \emph{TRACE} and EIT show structures extending up to
about 0.2 $R_{\odot}$ from the surface of the Sun, or about 1.5-3
coronal scale heights (at temperatures of 1-2 MK); this is surprising
because coronal volume emissivity in collisionally excited EUV lines
varies as the square of the density, hence the emissivity ratio
between the top and bottom of a hydrostatically supported active
region loop should be of order $e^{-3}-e^{-6},$ i.e. at most a few
times $10^{-2}$
(e.g. \citealt{Aschwanden2000,WinebargerWarrenMariska2003}), rather
than (as observed) of order $1$.  The tallest active region loops, at
~0.3 $R_{\odot}$, display yet more of a discrepancy.  A more careful
look at the geometry shows that the \emph{observed intensity} of an
unresolved coronal feature varies much more slowly than the
\emph{plasma emissivity}, resolving the discrepancy between the
hydrostatic scale height and observed intensity profile, without
dynamic or other support mechanisms to change the high altitude
emissivity.

EUV imaging in optically thin, collisionally excited lines such as
are visible through \emph{TRACE} measures a brightness integral along
the line of sight. The geometry of thin-feature imaging is summarized in Figure
\ref{fig:geometry}.  The brightness at any location in the image plane
of a non-vignetted EUV telescope may be described (\citealt{DeForest1991}\textbf{)}
by a brightness integral:\[
\Phi(x,y)=tf^{2}\int ds\left(K_{tel}(T(s,x,y))n_{e}^{2}(s,x,y)\right)\]
 where $\Phi$ is fluence (energy per unit area) on the detector plane;
$x$ and $y$ are focal plane coordinates; $t$ is exposure time;
$f$ is the f-ratio of the telescope; $s$ is distance along the line
of sight; $K_{tel}$ is a temperature response kernel that includes
the effects of telescope wavelength passband, atomic physics, and
the solar elemental abundances; and $n_{e}$ is local electron density
in the corona as a function of focal plane location and distance along
the line of sight. For DEM-style analyses, the integral is rewritten,
in the manner of Lebesgue, into an integral over T; but the simplest
case is a single, compact structure at one location along the line
of sight. 

Consider a telescope pixel with a line of sight that includes a single
elementary structure -- an isothermal thread that approximates a long
cylinder with slowly varying radius $r(s)$ along its length, with
uniform density across its cross section, and with no other structures
along the line of sight. If the thread is unresolved by the telescope,
and extends along the $x$ axis, then the fluence $\Phi$ deposited in each pixel 
is given by:
\begin{eqnarray}
\Phi_{ur}(x,y) & = & tf^{2}p\int\int dyds\left(K_{tel}(T_{struct})n_{e}^{2}\right)\nonumber \\
 &  & =tf^{2}p\pi r^{2}\, K_{tel}(T_{struct})n_{e,struct}^{2}\nonumber \\
 &  & \propto r^{2}n_{e,struct}^{2}\label{eq:brightness}
\end{eqnarray}
where $n_{e,struct}$ and $T_{struct}$ are the density and temperature
inside the structure, $p$ is the linear size of a pixel, and $r$
is the diameter of the (unresolved) solar structure being observed. 

Even if the structure is fully resolved and therefore the brightness
is spread across more pixels as the radius increases, geometry still
enters the brightness distribution as the depth of the feature varies.
In a resolved structure the brightness scales as
\begin{equation}
\Phi_{r}(x,y)\propto rn_{e,struct}^{2}\label{eq:brightness-2}
\end{equation}
under the same approximation as in Equation \ref{eq:brightness}.  
Clearly, structures near the resolution limit of a given telescope
will have a pixel brightness that varies between the 2nd and 1st
powers of $r$ as the geometry transitions between the fully unresolved
and fully resolved cases.

Consider, then, an unresolved coronal loop of temperature 
$10^{6}$K, height $0.2 R_{\odot}$ (2.7 scale heights) and a linear
expansion factor 8-10, typical of expansion ratios in modeled
active region loops (J. Klimchuk, priv. comm.). Then the pixel
brightness ratio $P_{top}/P_{base}$ is between $(8^{2})(e^{-5.4})$ and
$(10^{2})(e^{-5.4})$, or 0.3-0.5; this is consistent with brightness
ratios in active regions observed with \emph{TRACE}. A naive treatment, 
ignoring geometry, might predict a brightness ratio of just $e^{-5.4}$, or 
$4\times 10^{-3}$, which is much more contrast than is actually observed.

Hence, for unresolved, isothermal, elementary coronal structures that
expand with the bulk structure around them, we expect that
\begin{equation}
\frac{I_{ur}(z)}{I_{0}}\propto f_{local}^{2}(z)\, e^{-2z/h}\label{eq:I-ratio}
\end{equation}
where $z$ is height above the surface of the Sun, $I(z)$ is the
observed feature intensity in EUV, $I_{0}$ is its basal intensity,
$f_{local}$ is the local linear expansion ratio introduced for coronal
holes by \citet{MunroJackson1977}, and $h$ is the scale height (25-50
Mm for coronal plasma at 1-2 MK). 

Combining Equation \ref{eq:I-ratio} with a typical feature expansion
curve explains both the peculiar brightness of the EUV corona out to
about 1.2-1.3 $R_{\odot}$ and its sudden disappearance above that
height. Close to the surface, the expansion factor rises rapidly with
altitude, canceling the density gradient with altitude; but farther
away, the exponential falloff of density overwhelms the geometric
effect.  The balance is illustrated in Figure \ref{fig:Comparison of
exponential and parabolic growth}, which treats flux tubes in the
field from a potential dipole embedded 50 Mm under the surface of the
Sun; this field has a linear expansion ratio of 7.5 between the
photosphere and 0.2 $R_{\odot}$.

It is important to note that even threads that are resolved near the
top of an active region loop may not be resolved farther down. Such a
partially-resolved structure will follow the {}``unresolved''
intensity curves in Figure \ref{fig:Comparison of exponential and
parabolic growth} near its footpoints, and steepen gradually to be
parallel to the {}``resolved'' intensity curve as the cross section
exceeds the telescope's minimum resolvable width.  Both the
``resolved'' and ``unresolved'' curves fall off more slowly than the
simple $e^{-2z/h}$ curve that might be expected for a
constant-diameter feature, because of the increased depth at higher
altitudes.  Hence, all localized, optically thin structures that
expand with altitude (even those which are resolved spatially) will
have apparent intensity scale heights that are long compared to the
coronal density scale height, simply due to the geometry of the
structures.  This depth effect is important to image analysis of large
structures such as the diffuse brightening around active regions
(e.g. \citealt{Cirtain2005}), even though those structures may be
fully resolved in the image data.

\section{A hydrostatic active region model\label{sec:model}}

\begin{figure}[!tb]
\center{\includegraphics[%
  width=0.35\paperwidth]{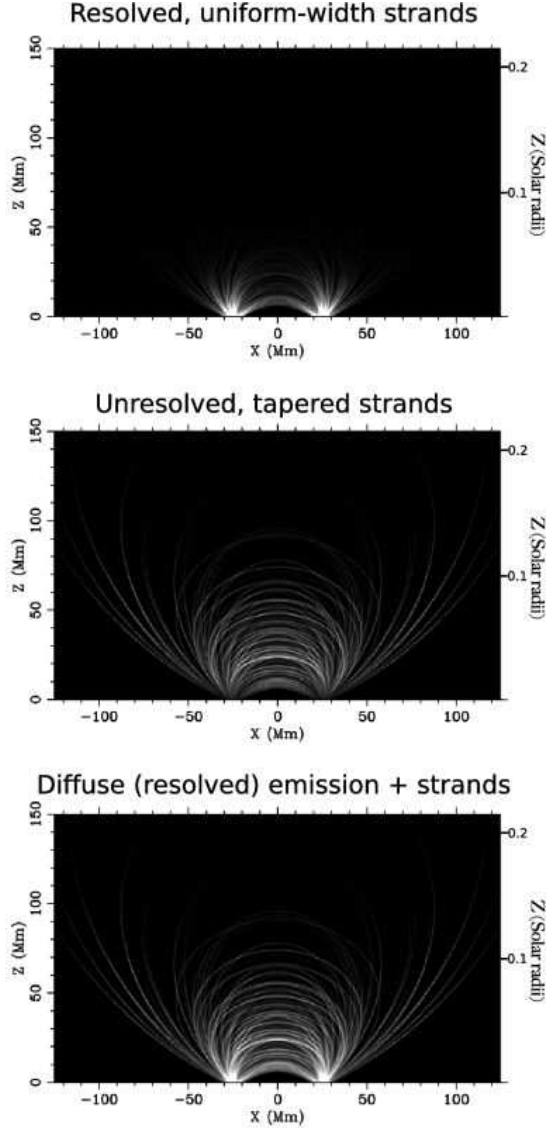}}

\caption{\label{fig:model}Rendered models of a hydrostatic active region:
(TOP) constant-width threads; (CENTER) unresolved variable-width threads.
(BOTTOM) A complete active region model including both diffuse (resolved)
emission and unresolved variable-width bright threads with low filling
factor.}
\end{figure}

To gain a better understanding of the simple analytic case above, I
generated hydrostatic coronal models of a very simple active region
by tracing a bundle of 300 random field lines through a potential
dipole field, and modeling collisionally excited, optically thin EUV
emission from a plasma structure associated with each of the field
lines. Each thread is modeled as a hydrostatic plasma with randomly
selected temperature between 1.0 - 1.5 MK and randomly selected base
density that varies over a factor of 10. The emission is calculated
using a simple broad-response-kernel approximation:
\begin{equation}
P_{pix}=\int d^{2}A_{pix}\left(\int ds\left(n_{e}^{2}\right)\right)\label{eq:model_emission}
\end{equation}
in which $P_{pix}$ is the optical power delivered to each pixel,
$d^{2}A_{pix}$ represents integration over the area of the pixel
in a perfect telescope, $ds$ is a coordinate along the line of sight,
and $n_{e}$ is the modeled electron density. This is similar to the
response-kernel method described above but neglects the $K_{tel}(T)$
temperature response kernel of the telescope (here approximated by
selecting only features in a temperature range appropriate to \emph{TRACE}
Fe XII images). For a collection of slowly-varying elementary structures,
and neglecting any background brightness, the triple integral over
$d^{2}A_{pix}ds$ reduces to a sum over all features along the line
of sight: 
\begin{equation}
P_{pix}=\sum_{i}V_{i,pix}n_{e,i}(z)\label{eq:sum}
\end{equation}
where $i$ indexes all the features in the model, $V_{i,pix}$ is the
volume occupied by the intersection of the pixel extended along its
line of sight and the $i^{th}$ feature of interest, and $n_{e,i}$ is
the density in the $i^{th}$ feature at the altitude of that
intersection.  The volume is determined by the length (measuring along
the feature in 3-space) of the intersection, and by the cross-section
of the feature.  The features are treated as small flux tubes, so
their cross sectional area is proportional to $1/B$, the reciprocal of
the local field strength.

The top two images in Figure \ref{fig:model} show the results
of the hydrostatic model for two feature geometries.  In the top frame,
the brightness at each pixel was calculated using a constant diameter
for each thread (assumed to have circular cross section). In
the center frame, the brightness was calculated using tapered
(but still unresolved) flux tubes. The models are projected into the
(X,Z) plane for the rendering, so that the height dependence can be
seen. The brightness of the two simple models corresponds well to
the plots in Figure \ref{fig:Comparison of exponential and parabolic growth}.

The \emph{apparent} increase in scale height of the threads in Figure
\ref{fig:model}(center) is only an optical illusion due to the small
filling factor and high density of the structures. The hydrostatic
scale height is approximately the same throughout the modeled corona
(varying by only a factor of 1.5), so that the emissivity of the
plasma in the threads must be very much brighter than that of the
surrounding material. The thread brightness is comparable to the
background brightness only because the underlying flux tube fills a
small percentage of each pixel; this attenuates the total brightness
of the thread, so that the underlying structure must be quite bright
to be visible at all. However, the $n_{e}^{2}$ dependence of
collisionally excited radiation accomplishes the job: to achieve an increase
of 100 in volume emissivity it is only necessary to increase density
(and hence pressure) by a factor of 10. In active region bases the
average value of $\beta$ is of order $10^{-4}$ or lower, which might
thus be raised to $10^{-3}$ in the dense structures.  Hence, the 
structures may still be magnetically dominated and yet also be dense
enough to be visible. The main practical limit to overall plasma
density in each structure is the total energy input, which must
balance the radiative losses along the entire length of the loop.

\begin{figure}[!tb]
\center{\includegraphics{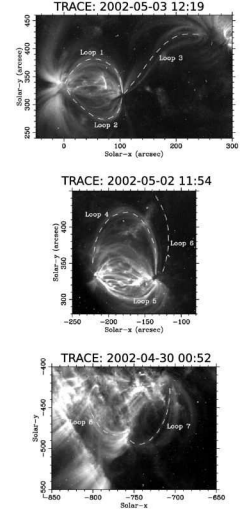}}

\caption{\label{fig:clips}Three active region images seen with \emph{TRACE}
in the 195 \AA\ EUV passband, selected for visual clarity of loop structures.
Several hand-traced loops composed of filamentary threads are marked.
}
\end{figure}

Put another way, the surprising aspect of Figure \ref{fig:model}(center)
is not that the tops of the thin structures are bright but rather
that their bases are faint. In this model, they are strongly attenuated
by the tapering geometry of the magnetic field.

Modeling only these super-bright unresolved elementary structures does
not yield a realistic image of a typical active region, because the
tapering of the structures produces footpoints that are too faint
(though some active region loops do indeed appear to have fainter
bases than tops in the EUV).  That problem may be addressed by
considering each active region to consist of a collection of very
dense, unresolved flux tubes embedded in a larger, resolved volume
with much lower density; this approach agrees with the experimental
result by \citet{Cirtain2005} that active regions have both fine and
diffuse components. The bottom panel of Figure \ref{fig:model}
demonstrates the result of summing the two types of emission, yielding
a fair approximation of a typical active region's brightness
structure with minimal physics (only potential-like magnetic expansion
factors and hydrostatic density profiles).  

One may expect that emission from the base of the active region will
be dominated by high-filling-factor plasma with close to the
conventional density and temperature values, but that visual features
at the top of the active region will be mainly small bright structures
with similar scale height and much greater density compared to the
bulk volume of the active region. In fact, the ratio of emission from
large-scale spaces and unresolved threads is expected to be more or
less the same at the top of the active region \emph{as a whole} and at
the bottom, but the expansion of the magnetic flux tubes between the
bottom and the top allows the threads to be distinguished spatially at
the top of the active region. At the base of the active region, both
the bright flux tubes and the spaces between them are smaller and
therefore superimposed by the telescope. There, the pixel brightness
is dominated by the material between the threads, because of the much
greater filling factor of the interthread material.

\section{Elementary Structures seen with \emph{TRACE?}\label{sec:trace}}

I studied the morphology of several active regions with threadlike
loop structure observed in the EUV with \emph{TRACE.} I present 8
typical cases in three separate active region complexes. The images
were selected for visual clarity of the loop structures in the active
region, and for variety of loop types. Each active region contained
bright threads that were isolated enough to afford visual
identification not only of the threads but also of complete
bundles. Three active region images, and some loop identifications,
are shown in Figure \ref{fig:clips}.

It is difficult to measure the taper of a curved coronal loop, so
I traced out individual bright features with a simple point-and-click
procedure, and then resampled the images to straighten the polygon by
splicing the piecewise-linear segments into a straight line segment
while retaining image scale. The resulting plots show the bright feature
and its environs as a straight line, much like a navigational chart
of a river. In these plots vertical is always a close approximation
of the transverse direction, so the width of the composite loop of
which each bright feature is a member is readily apparent.

\begin{figure*}[!t]
\includegraphics[%
  width=0.8\paperwidth,
  keepaspectratio]{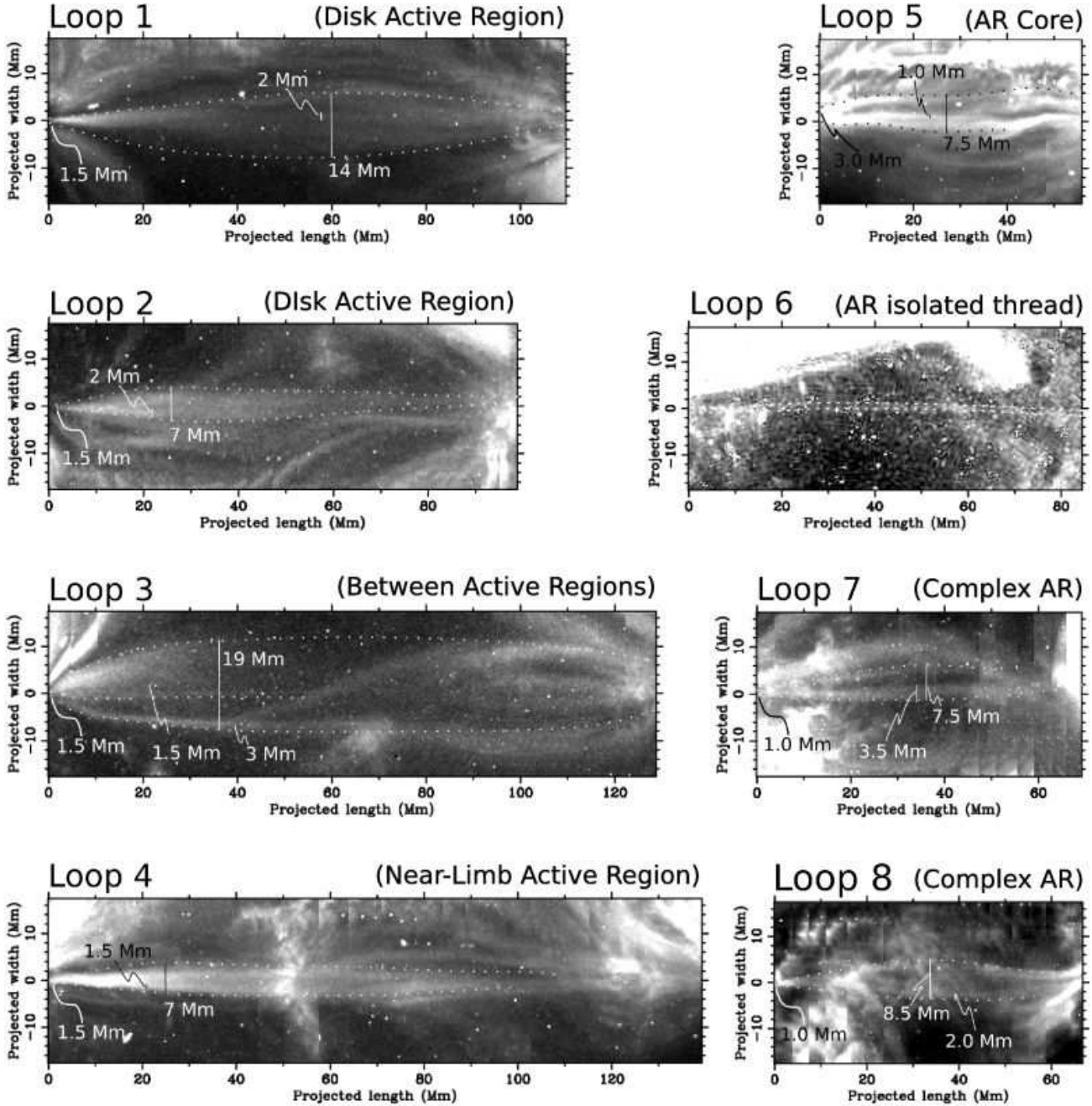}
\caption{\label{fig:linearized-segments}Loops from Figure \ref{fig:clips},
straightened along a hand-traced curve. Note the presence of small
constant-apparent-width features near the telescope resolution limit,
embedded within much larger variable-width loops. Tracings near the
left side of each image are more to be trusted than those near the
right. See text for full discussion.
}
\end{figure*}

I hand-traced the edges of the loop bundle containing each loop
thread, yielding the curved dashed lines in Figure
\ref{fig:linearized-segments}, and identified cross-sectional apparent
sizes of both the bundle and thread at several points. I selected
individual loops for clarity of features near one footpoint; that
footpoint is always at the leftmost side of the image, so that
tracings are more to be trusted on the left than on the right. That is
in keeping with Fuentes \& Klimchuk's \citeyearpar{Fuentes2006}
observation that features are difficult to trace
footpoint-to-footpoint. All of the loop bundles contain several bright
threads, each of which has similar appearance to the {}``elementary''
structures identified by \citet{Aschwanden2005b}; in fact, Loop 6 is
one of the structures identified by them as elementary. Note that
these constant-width threadlike features are at or near the
telescope resolution limit and are therefore not easily
distinguishable from much narrower structures that happen to not be
resolved; in fact, near the base of each loop bundle the several
threads appear to merge into a single bright feature with about the
same width as the individual threads themselves. Further, although
some of the features are clearly larger than the resolution limit near the
top of the loop (for example, Loop 7), if in fact they taper
with the surrounding bundle then each of them is much smaller than the
resolution limit near the footpoint; treating the features as if
they were resolved along their entire length thus results in a gross
underestimate of the expansion factor within the thread, even for
threads that are obviously resolved near their widest
points.

In each loop (other than Loop 6) in Figure
\ref{fig:linearized-segments}, the footpoint apparent width and the
loop central apparent width have been marked.  Without knowing the
unresolved profile of the features it is difficult to know the effect
of the \emph{TRACE} point spread function, because feature profile
strongly affects how the apparent width varies when a small feature is
convolved with a comparably sized kernel.  In a noise-free
measurement, feature widths add in quadrature, but the presence of
background beightness and the types of confusion demonstrated in
Figure \ref{fig:res_comparison} the noise-free FWHM of a feature may
be considerably different than its apparent visual width.

Taking the \emph{TRACE} point spread function to be 2.375 pixels (0.83
Mm) in width (the average of the minor and major PSF axes given by
\citet{Gburek2006}), and ignoring background or confusion effects in
the features, it is possible to generate rough estimates of the real
apparent widths of the structures that give rise to the features in
Figure \ref{fig:linearized-segments}.  Subtracting (in quadrature) the
Gburek width of 0.83 Mm gives expansion ratios varying from 2.6 (in
Loop 4) to 15.1 (in Loop 3), with a mean of 9.7 across all loops.
These numbers should be considered estimates only, because of the 
issues outlined in the previous paragraph.

Loop 3 in particular is interesting because it has both a high
expansion ratio and a faint thread with an apparent size of 1.5 Mm
(estimated real size of 1.25 Mm using subtraction in quadrature).  If we
assume that the cross-section of Loop 3 is self-similar, scaling
smoothly down to the footpoint, we find an estimate (which should be
taken as an upper bound) of 80km for the size of elementary structures
in the lower corona. Neglecting the point spread function entirely
yields a more conservative upper bound estimate of 110km: if the
underlying structure were larger at the base, it would appear wider
near the top of the active region.  We have thus derived an estimate
of the size of \emph{TRACE}-visible elementary structures at the base
of the corona, using hypothesis that the structures are barely
resolved by \emph{TRACE} near the loop top but taper proportionally
with their enclosing bundle as expected in a near-force-free plasma.

The 110km upper bound based on morphological taper can be corroborated
by the brightness profile of small, long, isolated features such as
that in Loop 6.  Such features (dubbed ``elementary'' by
\citet{Aschwanden2005b}) have constant apparent widths of
\textasciitilde{}1.5-2Mm, maximum altitudes of \textasciitilde{}0.1
$R_{\odot}$, and nearly constant brightness along their length. If indeed
these very long, very fine features are unresolved and supported
hydrostatically, then their footpoint sizes must be well under 100km
to sustain the brightness uniformity across their several scale
heights of altitude extent.

Using the fact that the structures are bright enough to be visible
with \emph{TRACE}, we can also derive a lower bound on their size.
The limiting minimum size of a particular observed structure is set by
the need to maintain $\beta\lesssim1$ near the footpoint of the
structures while emitting sufficient EUV photons per unit volume to be
visible against nearby resolved structures.  The smaller the
structure, the denser the plasma must be, with $\beta\sim1$
representing a practical maximum pressure (and density) for the
plasma.  

Taking $B=1000G$, $\beta=1$, and $T=10^{6}K$ at the base of
the corona over a sunspot, one finds a maximum electron density of
$10^{14}cm^{-3}$, compared to a {}``background'' diffuse density of
$10^{10}cm^{-3}-10^{11}cm^{-3}$ in the bases of active regions; this
sets the minimum size to a few $\times10^{-2}$ \emph{TRACE} pixels, or
\textasciitilde{}10 km, at the base of the active region. If the faint
structure at the center of Loop 3 (for example) were that narrow at
its base, it would be visible with about the correct brightness and an
actual width at its top of \textasciitilde{}150 km (0.5 \emph{TRACE}
pixel), which is consistent with the existing apparent width of 2
diagonal \emph{TRACE} pixels. Hence, elementary structures accessible
to detection by \emph{TRACE} most likely have a cross-field scale
between 10-100 km at the base of the corona. To fully resolve such
structures would require 15-150 milliarcsecond resolution from near
Earth, or 6 arcsecond resolution from a hypothetical solar probe
spacecraft located at 3-20 $R_{\odot}$.

One might not expect that many currently-visible features are much
smaller than 75km (the estimated size of Loop 3) at their bases:
Klimchuk (2006, priv. comm.) has pointed out that selecting the
brightest, clearest fine structures to study is equivalent to
selecting structures close to the resolution limit of the telescope.
This selection effect may bias the current result toward large
(\textasciitilde{}75km) filamentary structures.

\section{Discussion \& Conclusions\label{sec:discussion}}

I have demonstrated that spatial resolution effects are sufficient
to explain both the peculiar cross-sectional structure and extended
height of active region loops as seen with current EUV imagers, using
only hydrostatic equilibrium of the plasma contained in the loops
and geometric effects due to the non-resolved nature of the loops.
This offers simple explanations for several current theoretical difficulties
with observed active region loops, giving weight to the hypothesis
that elementary coronal structures are simply not resolved but are
affected by geometric effects that are not distinguishable to current
imagers.

The geometric effects attenuate the brightness of unresolved
structures near their bases. If ignored, this can result in a gross
underestimate of the basal density in the structures and hence an
overestimate of the pressure scale height within them. More generally,
semi-empirical analyses that compare the subjective appearance of
forward-modeled intensity data with solar images will yield incorrect
results if the geometry of unresolved structures is not incorporated
in the model. While some analyses (e.g.  \citealt{Fuentes2006})
consider spatial resolution issues, all current analysis of active
region loops seems to use the uniform apparent widths of narrow
coronal threads in \emph{TRACE} images as evidence of the uniformity
of the corresponding structures' actual widths. This inference gives
rise to many difficulties in the understanding of active region loops,
and it is arguably the weakest link in the current chain of inference
from observational results to comparison with theory.  Therefore the
question of loop width requires extremely careful consideration.

In particular, by comparing simultaneous images of a single feature
from both EIT and \emph{TRACE}, I have shown that resolution effects
can cause confusion in visual analysis even of structures up to
\textasciitilde{}6x the FWHM of the pixel-convolved PSF of an EUV
imaging instrument (EIT).  One may reasonably conclude that structures
with apparent sizes below about 6x the FWHM of the \emph{TRACE} PSF
(\textasciitilde{}13 \emph{TRACE} pixels) may also be subject to such
ambiguity and confusion. Simple analysis of the images themselves
cannot rule out such confusion effects, so that the sizes and even
unique identifications of features smaller than about 4-5 Mm across
are only weakly supported by image data from \emph{TRACE}.

It is important to understand that any difficulty with interpreting
\emph{TRACE} images of small features is not isolated to that
instrument: imaging distributed, optically thin objects is difficult,
and near the resolution limit of any telescope the interpretation of
the images becomes strongly model dependent.  Forward modeling of the
images produced by a particular type of structure is not sufficient
reason to conclude that the features observed in real data correspond
to resolved structures on the Sun.  Better resolution or, at a bare
minimum, truly adversarial hare-and-hounds type exercises are
required.

Furthermore, even fully resolving a coronal structure is not
sufficient reason to ignore geometric intensity effects. Even fully
resolved loops vary in thickness along the line of sight and that
variation must be considered and modeled in the course of drawing
inferences about scale height and other effects from image data.
Geometric effects in fully resolved structures are not as strong as in
unresolved structures, but are sufficiently important to feature
brightness profiles that they may act as a trap for the unwary.

This analysis is timely in part because much recent work attempts to
find physical mechanisms on the Sun for phenomena that could
potentially be understood in terms of instrumental effects.  I have
demonstrated that thread morphology in active region loops (and, by
extension, in similar structures such as quiet sun loops and polar
plumes) is not well constrained by current imaging instruments.
Taking possible resolution effects into account renders the imaging
data consistent with a naive hydrostatic model of the solar corona and
explains both the high feature contrast and relatively uniform height
(about $0.2-0.3 R_{\odot}$) of nearly all large, bright coronal
features seen with EUV imaging instruments.

Recent work by \citet{Aschwanden2005a} and \citet{Aschwanden2005b}
describes imaging of individual elementary loop structures in the
corona, based on differential emission measure analysis of individual
\emph{TRACE} images. Similar filter-ratio analyses
(e.g. \citealt{DeForest1995,Kankelborg1996}) have found
{}``elementary'' EUV structures (in the sense of being nearly
isothermal in multiple-passband analyses of EUV telescope data) that
were resolved by the \emph{Multi-Spectral Solar Telescope Array}
(\citet{Walker1991}) on spatial scales of \textasciitilde{}10
arcseconds, close to the resolution limit of that instrument; but it
is now obvious from the \emph{TRACE} data that multiple arcsecond size
structures are essentially always inhomogeneous.  

The present analysis suggests that recent results regarding coronal
elementary structures may be similar to the older ones: the structures
are likely not resolved by \emph{TRACE} in the usual sense.  I suggest
that, like earlier measurements, thread features in \emph{TRACE}
images are most likely distinguished merely by virtue of containing a
single bright structure (or group of structures at a particular
temperature) that happens to be much denser and brighter than other
adjacent magnetic structures passing through the same pixels in the
image plane.

What is different between the 1-arcsecond class data from \emph{TRACE}
and images from prior instruments is that \emph{TRACE} has sufficient
resolution to distinguish some of the bundled nature of coronal loops,
allowing the use of the taper of those bundles to infer something
about the fundamental scale of the corona. This was not possible with
images with multiple arcsecond resolution.

Further, the hypothesis that the small threads seen with \emph{TRACE}
are both elementary (isothermal and uniform density across the
structure) and unresolved also yields an estimate of the fundamental
size scale at the base of the corona based on the smallest features
seen higher up. That estimate (10km-100km) suggests that a high
resolution imager or solar probe mission will be needed to resolve
such elementary structures.  The upper size figure is derived (in
\S\ref{sec:trace}) from direct morphological scaling of observed
threads within active region loops, and the lower figure is derived
from brightness considerations and the need to confine the plasma
magnetically: elementary structures could in principle be smaller
still, but they would most likely be too faint to see with
\emph{TRACE}. The size range of 10-100 km is consistent with
reconnective heating induced by the motion of \emph{g-}band bright
points seen in the intergranular lanes of quiet sun and decaying
active regions, or by the motion of penumbral rolls and similar very
fine scale features near sunspots, suggesting that microflare
mechanisms driven by local surface motion may be responsible for the
large scale threaded appearance of active regions.

Taking the corona to have both a low-filling-factor, high
density component and a high-filling-factor, lower density component
(e.g. \citealt{Cirtain2005} and references therein) yields an elegant
explanation for the overall appearance and high feature contrast of
the EUV corona.  In particular, bright structures on the limb of full-disk images from
EIT and from \emph{TRACE} are easily seen to extend up to ~0.3 $R_{\odot}$ above
the surface of the Sun but fade rapidly into the background above that
altitude.  This effect may be seen as a direct result of the interplay
between expansion of fine structures and exponential decrease in their
density as in Figure \ref{fig:Comparison of exponential and parabolic
growth}.  

Finally, very dense, unresolved structures could account for the
surprisingly high intensity contrast of small features seen throughout
the corona with EUV imagers.  If bright coronal loops do indeed have
emissivities 100x-1000x higher than the surrounding ``diffuse''
corona, and are indeed tapered below the resolution limit of the
telescope, then the interplay of resolution and geometric effects
accounts very handily for two surprising aspects of EUV loops seen
with EIT, \emph{TRACE} and other EUV telescopes: the high intensity
contrast (order unity) of coronal loops at moderate altitudes of
0.1-0.3 $R_{\odot}$ compared to the background corona, despite a
factor-of-100 difference in length between the portion of the line of
sight inside the EIT loop and the portion in the surrounding corona;
and the comparatively low intensity contrast (of about unity) of
coronal loop footpoints compared to the surrounding corona, despite
large differences in altitude between the top and bottom of the loop.
The loops (and their components, \emph{TRACE} threads) may be
understood as very bright, unresolved structures that taper with
altitude, so that their integrated brightness close to the bottom of
the corona is quite small despite even higher emissivity than at
moderate altitudes of 0.2-0.3 $R_{\odot}$

Temporal behavior can be used to test the importance of geometric
effects and very dense threads, even without a high resolution
telescope.  The thermodynamics of long active region loops are
dominated by radiative cooling, with smaller contributions from
conduction and (in the presence of flow) advection.  Because the
emissivity of the plasma varies as $n_{e}^{2}$, the cooling time
scales inversely as density.  The radiative cooling time of typical
active region plasma is order of 1000 seconds, which is consistent
with the typical fading time reported by \citet{Schrijver1999} of ~500
seconds.  If threads are 10x denser than the surrounding material, the
cooling time should be correspondingly shorter - on the order of
perhaps 60 seconds.  This is not necessarily inconsistent if the
heating mechanism of the threads has a longer time scale and the
threads themselves are close to thermal equilibrium.  However, if no
such fast-fading threads are ever observed, that would weakly falsify
the proposition that coronal threads are currently-unresolved
filamentary structures.  Contrariwise, if even a small subset of
active region EUV threads are shown to fade with time scales much
faster than 5 minutes, that would support the proposition.

Throughout this discussion I have used the word {}``thread'' (and the
related phrases {}``threaded'', {}``multithreaded'', and
{}``threadlike'') rather than the more conventional {}``filamentary
structure''. The latter phrase is cumbersome and also leads to
confusion with filaments; while the more recent alternative,
{}``elementary structure'', has theoretical/modeling implications
unrelated to the simple appearance of the features. Similarly,
``loop'' is not specific to arcsecond scale features observed in high
resolution EUV images from instruments like \emph{TRACE}, because it
may also be considered to apply to larger bundles of threads, which
form 10-30 arcsecond scale features in moderate resolution EUV images.
I suggest using {}``thread'' as a purely observational term to
describe small, apparently-constant-width structures within coronal
loops and other features in the image plane of a telescope, because
the word is short, pithy, easy to remember via analogy to textiles in
everyday experience, and not (yet) laden with non-observational
nuance.

\acknowledgements{The author thanks the \emph{TRACE} and EIT teams for making
their data readily available, T. Tarbell of Lockheed for helpful
discussion of the \emph{TRACE} spatial resolution, J. Klimchuk of the Naval
Research Laboratory for much illuminating discussion of active
regions, S. McIntosh and M. Wills-Davey of the Southwest Research
Institute for comments and input on structure and clarity, and the 
anonymous referee for his great patience and invaluable help.  SOHO is a
project of international collaboration between NASA and ESA.  Data
analysis and modeling were performed with the Perl Data Language,
which is freely available from http://pdl.perl.org. This work was
funded in part by NASA's SR\&T/SHP program and in part by the
Southwest Research Institute.}

\bibliographystyle{plainnat}


%
\clearpage
\clearpage


%
\clearpage
\clearpage


%
\clearpage
\clearpage


%
\clearpage
\clearpage


%
\clearpage
\clearpage


%
\clearpage
\clearpage


%
\clearpage
\clearpage


%
\clearpage
\clearpage

\end{document}